\def\daga#1{{#1\mkern -10.0mu /}}
\def\'#1{\if#1i{\accent19\i}\else{\accent19#1}\fi}
\newcommand{\be}{\begin{eqnarray}}
\newcommand{\ee}{\end{eqnarray}}
\documentstyle[preprint,aps]{revtex}
\begin{document}
\draft
\title{Bound on  the   neutrino  magnetic moment from  chirality flip in supernovae}
\author{Alejandro Ayala and Juan Carlos D'Olivo}
\address{Instituto de Ciencias Nucleares\\
         Universidad Nacional Aut\'onoma de M\'exico\\
         Aptdo. Postal 70-543. M\'exico D.F. 04510, M\'exico.}
\author{Manuel Torres}
\address{Instituto de F\'{\i}sica\\
         Universidad Nacional Aut\'onoma de M\'exico\\
         Aptdo. Postal 20-364, M\'exico D.F. 01000, M\'exico.}
\maketitle
\begin{abstract}

For neutrinos with a magnetic moment, we show that the collisions in a  hot and dense 
plasma  act as an efficient  mechanism  for  the conversion of $\nu_L$ into $\nu_R$.   
The production rate for right-handed neutrinos is computed in terms of
a resummed photon propagator which consistently incorporates the background 
effects. Assuming that the entire energy in a supernova collapse is not 
carried away by the $\nu_R$, our results can be used to place an upper limit 
on the neutrino magnetic moment
$\mu_\nu < (0.1-0.4)\times 10^{-11}\mu_B$.

\end{abstract}

The properties of neutrinos have become the subject of
an increasing research effort over the last years. Amongst these properties,
the neutrino magnetic moment $\mu_\nu$ has received attention in connection
with various chirality-flipping processes that could have important consequences
for the explanation of the solar neutrino problem \cite{cisne,volo1}, and for
the dynamics of stellar collapse   \cite{nussi,barbi}.
As a consequence of a non-vanishing magnetic moment, 
left-handed neutrinos produced  inside the supernova core during the collapse,
could change their chirality becoming sterile with respect to the weak interaction. 
These sterile neutrinos would fly away from the star leaving essentially  
no energy to explain the observed luminosity of the supernova.
The chirality flip could be caused by the interaction with an external magnetic
field or by the scattering with charged fermions in the background for instance
$\nu_L \, e^{-}  \to \nu_R\, e^{-}$ and $\nu_L \,  p  \to \nu_R\, p$. 
Invoking this last mechanism, Barbieri and Mohapatra~\cite{barbi} have
derived a limit $\mu_\nu < (0.2-0.8)\times 10^{-11} \mu_B$.

Dispersion processes in a plasma could exhibit infrared divergences due to the 
long-range electromagnetic interactions. To prevent such divergences,
the authors in Ref.~\cite{barbi} introduced an ad hoc thermal mass into the 
vacuum photon propagator. 
However, it is well known that at high temperature, a consistent formalism 
developed by Braaten and Pisarski~\cite{brat1,LeBellac},
rendering gauge independent results, requires the use of effective 
propagators and vertices that resum the leading-temperature corrections.
In particular, this method has been applied to the study of the damping rates 
and  the energy loss of particles propagating through hot 
plasmas~\cite{brat2,brat3,brat4}.
In this paper we show that the above formalism provides also a convenient 
framework to study the neutrino chirality flip
in a dense plasma \cite{rafn1}.
 
The spectral function of the resummed photon propagator in a plasma  presents a cut for
space-like momenta; the physical origin of this purely thermal cut  is 
Landau damping \cite{LeBellac}. In this case, the conversion from 
left-handed to right-handed neutrinos happens through scattering by the 
exchange of effective space-like photons.  We compute the production rate of 
$\nu_R$'s and the corresponding luminosity for such a process in a supernova.  
We recall that the contribution of the plasmon decay 
$\gamma \to \bar{\nu}_L \nu_R$ to $\nu_R$
production is less important  than  the chirality flipping process due to the 
high neutrino density in the supernova core.
Our result can be used to place an upper bound on the neutrino magnetic 
moment which is in the range $\mu_\nu < (0.1-0.4)\times 10^{-11}\mu_B$,
where $\mu_B$ is the Bohr magneton.

Consider a QED plasma in thermal equilibrium at a temperature $T$ and 
with an electron chemical potential $\tilde{\mu}_e$ such that 
$T\, , \tilde{\mu}_e \gg m_e$, where $m_e$ is the mass of the electron.
The production rate $\Gamma$ of a right-handed neutrino with total energy 
$E$ and momentum  $\vec{p}$ can be conveniently expressed in terms of the 
$\nu_R$ self-energy $\Sigma$ as~\cite{weldon}
\be
\Gamma\left(E\right)= \, \frac{1}{2E} \,  n_F \, \,
 {\mbox T}
   {\mbox r}\, \left[ \daga{P} \, R \,   {\mbox I}{\mbox m}\,  \Sigma
   \, \right]   \label{rate}\, ,
\ee
where, $L,R = {1\over 2} \left(1 \pm \gamma_5\right)$ and  $n_F$ is the Fermi
distribution for the  $\nu_R$'s. Since the $\nu_R$'s are sterile, we could set 
$n_F=1$ from the beginning, however as we shall see the $\nu_R$ distribution
cancels in the final result. In the above expression,
Im$\Sigma$ can be directly computed following either the
imaginary or the real-time formulations of thermal field theory with identical 
results \cite{nos}. In what follows we will work in the imaginary-time
formalism.

The one loop contribution to $\Sigma$  is given 
explicitly by
\be
   \Sigma(P)= \mu_\nu^2T\sum_{n}\int\frac{d^3k}{(2\pi)^3}
  \,  K_\alpha \,  \sigma^{\alpha\rho}\,  S_F \left( \daga{P}  \, - \,    {K
\mkern -13.0mu / } \right)
     \, L  \,   K_\beta \,  \sigma^{\beta\lambda} \,\, ^*D_{\rho\lambda}(K)
   \label{self} \, ,
\ee
where $P = (i p_0, \vec p)$ and $K = (i k_0, \vec k)$ are the 
four-momenta of the incoming neutrino and the virtual photon, respectively
with $p_0= (2m+1)\pi T$ and $k_0 =  2n \, \pi T$, $m$ and $n$ 
being integers; and $p \equiv |\vec p|$, $k \equiv |\vec k|$.
 In the integration region where the momentum
$k$ flowing through the photon line is soft 
($i.e.$ of order $eT$), hard thermal
loop (HTL) corrections to the  photon propagator contribute at  leading order
in $e$ and must be  resummed. The effective propagator  is obtained by summing the
  geometric series of
one-loop self-energy  corrections proportional to $e^2 T^2$.
The intermediate neutrino line can  be taken  as a bare   fermion propagator
$S_0$, because the  $\nu_L$ propagator gets dressed only through weak
interactions  with the particles in the medium. For the neutrino-photon vertex
we use  the magnetic dipole  interaction $\mu_\nu  \sigma_{\alpha \beta}
K^{\beta}$. The neutrino effective electromagnetic vertices are of no concern 
to us here since they are induced by the weak interaction of the charged 
particles in the background and thus conserve chirality.

In a covariant gauge the HTL approximation to the photon propagator 
$ ^*D^{\mu\nu}$ is
\be
^*D^{\mu\nu}(K) &=&  ^*\Delta_L(K) \, {P^{\mu\nu}_L} \, + \,
^*\Delta_T(K) \, {P^{\mu\nu}_T}   \label{prop} \, ,
\ee
where $P^{\mu\nu}_L$, $P^{\mu\nu}_T$
are the  longitudinal and transverse projectors, respectively.
We drop the term proportional to the gauge parameter since it does
not contribute to $\Sigma$, as can be easily checked.  The scalar functions
$^*\Delta_{L,T}(K)$ are given by
\be
 ^*\Delta_L(K)^{-1} \, &=& \,  K^2 + 2m_\gamma^2 \,\frac{K^2}{k^2}
\left[1- \left(\frac{i k_0}{k} \right)
   Q_0 \left(\frac{i k_0}{k}\right)\right] \, ,   \nonumber \\
    ^*\Delta_T(K)^{-1} \, &=&
  - K^2 -  m_\gamma^2 \, \left(\frac{i  k_0}{k} \right)\left[
   \left[1- \left(\frac{i  k_0}{k} \right)^2 \right]Q_0\left(\frac{i
k_0}{k}\right)
   + \left(\frac{i k_0}{k}\right)\right]
  \label{deltas} \, ,
\ee
where $Q_0(x)=\frac{1}{2}\ln\left(\frac{x+1}{x-1}\right)$ is a Legendre function
of the second kind and $m_\gamma$ is the photon thermal mass.
In the limit $T,\tilde{\mu}_e \gg m_e$,
\be
   m_\gamma^2=\frac{e^2}{2 \pi^2}\left(\tilde{\mu}_e^2 +
       \frac{\pi^2 T^2}{3}\right)
   \label{mass} \, .
\ee

The sum over loop frequencies in Eq.~(\ref{self}) is evaluated by 
introducing the spectral representation for the propagators and summing over 
$n$.  At the end the neutrino energy is analytically continued to the real 
Minkowski energy $i p_0 \to E + i \epsilon$. The spectral representation
of the fermion propagator is~\cite{vija}
\be
S_0(p_0,p) \, = \,  \int^{\beta}_{0} d\tau  e^{i p_0\tau}
\left[  P_+ \, \tilde{f}(- E) \,e^{- E\tau} \,  + \,
P_ -\tilde{f}(E ) \, e^{E\tau}  \right]
\label{rspf}  \, ,
\ee
where $\tilde{f}(z)=  (e^{(z-\tilde{\mu}_\nu)/T} +1)^{-1}$ denotes the
Fermi-Dirac distribution   with  $\tilde{\mu}_\nu$ the $\nu_L$ chemical
potential and  $P_\pm = \left(E \gamma^0  \pm i \vec{p} \cdot \vec{\gamma} \right)/(2 E)$ 
are the positive and negative energy projection 
operators for massless particles.

 For the electromagnetic fields the spectral representation
reads
\be
  ^*\Delta_{L,T}(k_0,k) \, = \,  \int^{\beta}_{0} d\tau  e^{i k_0\tau}
\int_{-\infty}^{\infty} d \omega  \, \,  \rho_{L,T} (\omega, k) \, \left[
1 + f(\omega)\right] \, e^{- \omega \tau}
   \label{rspb} \, ,
\ee
where  $f(z)=(e^{z/T}-1)^{-1}$ is the Bose-Einstein distribution. The 
spectral densities
$\rho _{L,T}(\omega\,,k)$ are obtained from the imaginary part of
$^* \Delta(k^0, k)$ after analytical continuation
$ \rho_{L,T}(\omega\,,k) = 2\,{\mbox I}{\mbox m}
   \Delta_{L,T}\left({i k_0\rightarrow \omega+i\epsilon},k\right) $, where 
the $+,-$ signs correspond to the $L,T$ modes respectively. The spectral densities $\rho_{L,T}(\omega\,,k)$  contain the discontinuities
of the photon propagator across the real-$\omega$ axis. Their support
depends on the magnitude
of the ratio between $\omega$ and $k$. For $\left| \omega/k \right| > 1$,
$\rho_{L,T}(\omega\,,k)$ have support on the points $\pm \omega_{L,T}(k)$,
i.e., the time-like quasiparticle poles. In the space-like region the support of 
$\rho_{L,T}(\omega\,k)$ lies on  the whole interval $-k<\omega<k$, with the
contribution arising from the branch cut of $Q_0$.
In evaluating the self-energy for the chirality  flip process $\nu_L \to \nu_R$,
 the kinematically allowed region is
restricted to space-like momenta $|\omega|  \leq k$.
In this case from Eqs.~(\ref{deltas})  we obtain
\be
   \rho_L(\omega\,,k)=   \,  \frac{x}{(1 - x^2)} \,
   \frac{2 \pi m_\gamma^2 \,  \theta\left( k^2 - \omega^2\right) }
   {\left[k^2+2m_\gamma^2 \, (1-\frac{x}{2}\ln\left|\frac{x+1}{x-1}\right|)
   \right]^2+\left[\pi m_\gamma^2\,  x\right]^2}\, ,
   \label{ro1}
\ee
\be
   \rho_T(\omega\,,k)= \,
 \frac{  \pi  m_\gamma^2 \, x \, \left( 1 - x^2  \right) \, \theta \left(k^2 - \omega^2\right)}
  {\left[ k^2(1 -x^2)+m_\gamma^2 \, \left( x^2+
    \frac{x}{2}(1 - x^2)\ln\left|
  \frac{x+1 }{x-1}\right|\right)\right]^2 +
   \left[\frac{\pi}{2} m_\gamma^2\,  x \left(1 - x^2\right)\right]^2} \, .
   \label{ro2}
\ee
where we have defined $x\equiv \omega/k$.

 The rate of production  of right-handed neutrinos in Eq. (\ref{rate}) can be computed after 
the substitution of the   spectral representations (\ref{rspf}) and  (\ref{rspb}) into (\ref{self}),
our final result is 
\be
   \Gamma(E) &=& \frac{\mu_\nu^2}{32 \pi^2  E^2}\int_0^{\infty}k^3 dk \,
\int_{-k}^{k}d\omega
\,   \theta(2E+\omega-k)   \left(1 + f(\omega)  \right) \, \tilde{f}_\nu(E + \omega
)\nonumber\\
   && (2E+\omega)^2 \,   \left(1 - \left(\frac{\omega}{k} \right)^2 \right)^2  \, \,
   \left[\rho_L(\omega,k)+
   \left( 1 - \frac{k^2}{(2E+\omega)^2}\right)\rho_T(\omega,k)\right]\, .
   \label{game}
\ee
Both, longitudinal and transverse photons, contribute to this rate.
Notice that in addition to $|\omega| \leq  k$ there is also
the restriction $(k -\omega )\leq 2E$ coming from the condition 
$|\cos\Theta|\leq 1$, where $\Theta$ is the angle between the momenta of the 
incoming neutrino and the virtual photon. With these restrictions
the  integrand in  the previous equation can be proved to be positive definite.
The rate  $\Gamma$ can be written as the sum of two contributions
$\Gamma = \Gamma_e + \Gamma_a$, that correspond to the production of $\nu_R$
through the emission or absorption of a virtual photon.
$\Gamma_e$  comes from the interval $0\leq \omega < k$, whereas $\Gamma_a$
corresponds to the interval $-k < \omega \leq 0$, as can be checked by means of 
the identity $1+f(\omega)+f(-\omega)=0 $ and the substitution $\omega\rightarrow -\omega$ 
in this second interval.

In order to illustrate our results we  consider the  emission of right-handed
neutrinos immediately after a supernova core collapse.
The large mean free path of the right handed neutrinos compared to the core 
radius implies that the $\nu_R$'s would freely fly away from the supernova.
Therefore, the core luminosity for $\nu_R$ emission can be computed as
\be
Q_{\nu_R}= V \int_0^{\infty} {d^3 p \over \left( 2\pi \right)^3}  \, E \,
\Gamma(E)
    \label{lumi}  \, ,
\ee
where $V$ is the plasma volume and $E= p$.
To make a numerical estimate, we shall  adopt a simplified  picture of the 
inner core, corresponding to the the average parameters of SN1987A 
\cite{mohapatra,burro}. Consequently, we take a constant density 
$ \rho \approx 8 \times 10 ^{14} \, {\rm g/cm}^3$, a volume
$V \approx  8 \times 10^{18} \,  {\rm cm}^3$, an electron to baryon
ratio $Y_e \simeq Y_p \simeq 0.3$, and   temperatures in the range
$T = 30 \sim 60 \, {\rm MeV}$.
This corresponds to a degenerate electron gas with a chemical potential
$ \tilde{\mu}_e$ ranging from $307$ to $280 \,  {\rm MeV}$.
For the left-handed neutrino we take $\tilde{\mu}_\nu \approx 160 \,
{\rm MeV}$.
Using this values in  Eqs.~(\ref{game}) and (\ref{lumi}), we obtain by
numerical integration
\be
 Q_{\nu_R} = \left( { \mu_\nu \over \mu_B  }  \right)^2  \,(0.7-4.3)\times
10^{76}\,
{\mbox e}{\mbox r}{\mbox g}{\mbox s}/{\mbox s} \,  ,
   \label{numlum}
\ee
for $T$ ranging from $30$ to $60$ MeV. 

Assuming that the   $\nu_R$ emission lasts  for about   one second,  the luminosity
bound is $Q_{\nu_R}\leq 10^{53}$ ergs/s which places the upper limit
on the neutrino magnetic moment
\be
 \mu_\nu < (0.1-0.4) \times 10^{-11}\mu_B\, .
   \label{momm}
\ee

This upper bound slightly improves the result previously obtained by Barbieri 
and Mohapatra~\cite{barbi}. As mentioned before, these authors consider
the helicity flip scattering $\nu_L e \to \nu_R  e$  to order $e^2$ introducing the Debye
mass in the photon propagator as an infrared regulator. Although  the integrated luminosity shows no important dependence on the use of the complete leading order perturbation theory, the  $\nu_R$ spectrum  does, as we discuss below. 

 The usual bare perturbation  theory to order $(e\mu_\nu)^2$  can be recovered \cite{LeBellac} by simply neglecting  $m_\gamma^2$ in the denominators in  Eqs.  (\ref{ro1}) and (\ref{ro2}).
 This leads to infrared divergent results. The usual 
screening prescription consists of  cutting  off these infrared  divergences  treating both the longitudinal and transverse photons as  particles of mass $\omega_p$; with the plasma frequency $\omega_p^2 = (2/3) m^2_\gamma$. 
The  corresponding approximation for  the  spectral densities is 
 \be
\rho_L \approx  2\rho_T \approx   {2 \pi m_\gamma^2 \omega /k^3  \over  
k^2  -  \omega^2   + \omega_p^2  } 
\label{aprox} \, , 
\ee
instead of using  the expressions  in  (\ref{ro1}) and (\ref{ro2}).   Fig. 1 shows  the rate of  production of right-handed neutrinos,  for  $T = 30 MeV$ and
 $ \tilde{\mu}_e = 307 MeV$,   as a function  of  their   energy.  
We notice that for soft neutrino energies 
($E {\ \lower-1.2pt\vbox{\hbox{\rlap{$<$}\lower5pt\vbox{\hbox{$\sim$}}}}\ } e T$)  the
use of formula  (\ref{aprox}) underestimate the  contribution 
by several  orders of magnitude as compared to the complete  leading-order  result. 
On the other hand,   as the energy increases above the hard  scale $\sim T$,  the two approaches  lead to very similar results. 
Let us remark that the approximation  in  Eq.(\ref{aprox}) correctly 
reproduces the  static limit of the longitudinal propagator:
$\Delta_L (0, k)^{-1} = k^2 + \omega_p^2 $,
but it   also predicts a similar behavior for $\Delta_T (0, k)^{-1} $, and thus 
it is not in agreement with the 
absence of magnetic screening in the static limit for the transverse mode.
The vanishing  of  the  magnetic mass in the HTL approximation of the photon propagator
leads  to 
infrared singularities in certain  quantities \cite{blaizot}.
In our  case, there are enough powers of $k$ coming from the vertex factors in (\ref{self}) to
render the $\nu_R$ production rate finite.

\vskip1.0cm

\let\picnaturalsize=N
\def\picsize{4.0in}
\def\picfilename{figu1.eps}
\ifx\nopictures Y\else{\ifx\epsfloaded Y\else\input epsf \fi
\let\epsfloaded=Y
\centerline{\ifx\picnaturalsize N\epsfxsize \picsize\fi \epsfbox{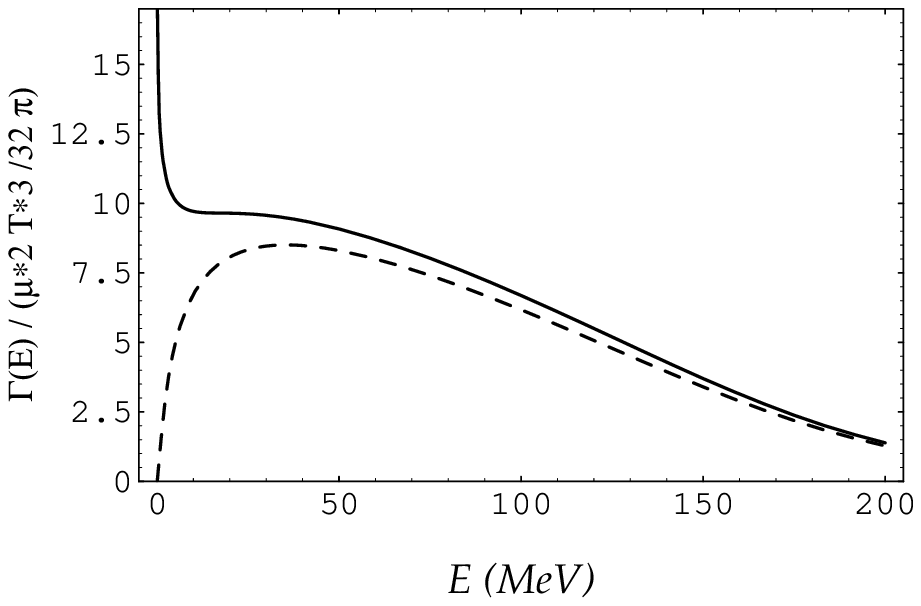}}}\fi

{\it  Fig. 1. The $\nu_R$ production rate as a function of the energy.  The complete
 leading-order  result (solid-line) is compared with the calculation in which the infrared 
divergence is cut off by  an $ad$ $hoc$   prescription (dotted-line)
in which  both the longitudinal and transverse photons are considered  as  particles of mass $\omega_p$. The parameters are selected as: 
$T = 30$ MeV, $\tilde{\mu_e} = 307$ MeV and $\tilde{\mu_\nu} = 160$ MeV.  }

The use of the screening prescription  in Eq.(\ref{aprox})  also underestimates the contribution to the luminosity. However, the  phase space factor in 
Eq.~(\ref{lumi}) suppresses the contribution to the luminosity for small values 
of $E$. An explicit calculation along this line
yields a luminosity that  varies between $(3-5)\%$  from
our result in Eq.~(\ref{numlum}). We conclude that  the  $\nu_R$ production
spectrum  for soft $E$  is sensitive  to the form of the resummed photon
propagators. Yet, the  luminosity  depends very little  on these effects,
provided  an infrared convergent factor of the correct magnitude is 
introduced. A more detailed analysis will be presented elsewhere~\cite{nos}.

A word of caution should be mentioned in relation to  the   result  in
Eq.~(\ref{momm}). It has been pointed out by Voloshin~\cite{volo2} that the
$\nu_R$'s produced by the magnetic moment interaction could undergo resonant
conversion back into $\nu_L$'s through spin rotation  in the magnetic field of
the supernova core, with the subsequent trapping of the $\nu_L$'s  by the 
external layers. If this is the case, then the  bound  in Eq.~(\ref{momm}) 
becomes meaningless. However, the core density is rather high and the matter 
effect might dominate over the $\mu_\nu\,B$ term, suppressing the flip back of 
$\nu_R$ to $\nu_L$~\cite{mohapatra}.

Recently, another mechanism for the neutrino chirality flip has been proposed,
which occurs via the \v{C}erenkov emission or absorption of plasmons in
the supernova core~\cite{mohanty}. Since the photon dispersion relation in 
a relativistic plasma shows a space-like branch for the longitudinal mode,
the \v{C}erenkov radiation of the plasmon is, in principle, kinematically 
allowed~\cite{jc}. However, this mode develops a
large imaginary  part, which implies that the Landau damping mechanism
acts to preclude its propagation as we have discussed. Consequently, we 
think that no better than the quoted limit in Eq.~(\ref{momm}) can be derived 
by this type of neutrino chirality flipping processes in a supernova core.
Let us notice  that Eq.~(\ref{momm}) is comparable to  a reliable constraint, 
$ \mu_\nu < 0.2 \times 10^{-11}\mu_B $, 
that  has been  derived by Raffelt  \cite{rafn2}  from   the analysis of plasmon decay 
in  globular-custer stars.

To conclude, we have shown that the collision processes in a hot and dense
plasma, allow
for the efficient conversion of  $\nu_L$  into  $\nu_R$.
In this work, plasma effects are consistently taken into account by
means of the resummation method of Braaten and Pisarski. For soft values of
the energy, the production rate for $\nu_R$'s differs significantly  from that 
obtained by a constant Debye mass screening prescription.  However,
correction to the integrated luminosity are small.
For this reason, our upper bound on the neutrino magnetic moment 
does not differ significantly from the one obtained from the cooling of SN1987
by Barbieri and Mohapatra~\cite{barbi}. Knowledge of an accurate
expression for the $\nu_R$ production rate, as given in Eq.~(\ref{game})
could be of importance in a detailed analysis of supernova processes.

\acknowledgments

This  work  was  supported  in part  by  the Universidad Nacional Aut\'onoma de
M\'exico under  Grants  DGAPA-IN100694 and DGAPA- IN10389, and by
CONACyT-M\'exico under  Grants  3097 p-E and I27212-E.

\end{document}